\documentclass[12pt]{article}
\usepackage[dvips]{graphicx,color}
\usepackage[latin1]{inputenc}
\usepackage{amsmath}
\usepackage{amsfonts}
\usepackage{amssymb}
\newcommand{\be}{\begin{equation}}
\newcommand{\ee}{\end{equation}}
\newcommand{\ben}{\begin{eqnarray}}
\newcommand{\een}{\end{eqnarray}}
\title{Vacuum polarization in $d+\frac{1}{2}$ dimensions}
\author{C.~D.~Fosco$^{a}$\\
A.~P.~C.~Malbouisson and I.~Roditi$^{b}$\\
{\normalsize\it $^a$Centro At\'omico Bariloche and Instituto Balseiro}\\
{\normalsize\it Comisi\'on Nacional de Energ\'\i a At\'omica}\\
{\normalsize\it 8400 Bariloche, Argentina}\\
{\normalsize\it $^b$Centro Brasileiro de Pesquisas F\'{\i}sicas}\\
{\normalsize\it Rua Dr. Xavier Sigaud 150,}
{\normalsize\it Rio de Janeiro, RJ, Brazil}
}
\begin{document}
%
\date{}
\maketitle
%
\begin{abstract}
\noindent We study the main properties of the one-loop vacuum polarization
function ($\Pi_{\alpha \beta}$) for spinor $QED$ in `$d + \frac{1}{2}$
dimensions', i.e., with fields defined on ${\mathcal M} \subset
{\mathbb R}^{d+1}$ such that ${\mathcal M} = \{(x_0,\ldots,x_d) |
x_{d}\geq 0 \}$, with bag-like boundary conditions on the boundary
$\partial{\mathcal M} = \{(x_0,\ldots,x_d) | x_{d}= 0 \}$.  We
obtain an exact expression for the induced current due to an external
constant electric field normal to the boundary.  We show that, for the
particular case of $2+1$ dimensions, there is a transverse component
for the induced current, which is localized on a region close to
$\partial{\mathcal M}$. This current is a parity breaking effect
purely due to the boundary.
\end{abstract}
\maketitle
There are many interesting quantum field theory models where the
presence of boundaries is relevant to the physical properties of the
system. Noteworthy examples are the Casimir
effect~\cite{Casimir:1948dh,Bordag:2001qi}, the bag model of
QCD~\cite{Chodos:1974pn}, as well as many condensed matter physics
models~\cite{Fuentes:1995ym}.

One of the most interesting consequences of the presence of boundaries
in a quantum field theory is that they strongly influence the
structure of the quantum fluctuations.  Besides, response functions
are determined by the correlation between fluctuations; therefore,
those functions should also be quite sensitive to the presence of
boundaries. In particular, one should expect a strong effect on the
$QED$ vacuum polarization function, since a bag-like condition implies
that the normal component of the current due to an external field
vanishes on the boundary. Since that current can be expressed in terms
of the vacuum polarization function, the latter should depend strongly
on the distance between its spatial arguments and $\partial{\mathcal M}$.  In
other words, the conductivity tensor becomes space-dependent and also
anisotropic in the presence of a bag-like condition.

In this letter, we calculate the vacuum polarization function for
massless fermions in a $d+1$-dimensional region ${\mathcal M}$,
defined by the conditions $x_d \geq
0$~\footnote{Following~\cite{Fuentes:1995ym} we use the expression
  `$d+\frac{1}{2}$ dimensions' to denote such a spacetime.}. Equipped
with the general expression for $\Pi_{\alpha \beta}$, we then
derive the induced current when an external constant electric
field is applied in the direction normal to the `wall' at $x_d =
0$. We shall show that, for the case $d=2$, there is an induced
current in the $x_1$ direction when a constant electric field in
the $x_2$ direction (i.e., normal to the wall) is applied. This is
a parity-breaking effect due to the fact that the spatial region
itself is not invariant under parity transformations. We recall
that, in $2+1$ dimensions, those transformations are really
reflections about an axis~\cite{Deser:1981wh}.

The calculation of $\Pi_{\alpha\beta}$ in the standard perturbation theory
approach (which we shall follow) requires the knowledge of the fermion
propagator in a region with boundaries. To that end, the multiple
reflection expansion (MRE)~\cite{BB1,BB2,HJ1,HJ2} provides a
systematic approach to the calculation of quantum field theory
propagators in a spacetime region ${\mathcal M}$, with a non-trivial
boundary ${\mathcal S} \equiv \partial {\mathcal M}$.  The different terms in the
MRE are constructed with the standard (no-boundary) free propagator,
and can be conveniently ordered according to the increasing number of
`interactions' with ${\mathcal S}$.

$S_F$ is determined by the following equations:
\begin{equation}\label{eq:deffer}
\left\{ \begin{array}{lcl}
( \not \! \partial + m ) S_F(x,y) \;=\; \delta^{(d+1)} (x-y) \;\;&,&\;\;
\forall x, y
\in {\mathcal M} \\
\lim_{x\to\alpha} P_\alpha \, S_F(x,y) \;=\; 0 \;\;&,&\;\;
\forall \alpha \in {\mathcal S}\;,
\;\; \forall y \in {\mathcal M} \;\;,
\end{array}
\right.
\end{equation}
where $P_\alpha \equiv \frac{1 + \gamma \cdot n(\alpha)}{2}$,
$n(\alpha)$ is the (outer) normal at a given point $\alpha \in
{\mathcal S}$, and the $x \to \alpha$ limit should be taken {\em
from inside\/} ${\mathcal M}$.

 We shall assume that the fermions are massless. In this case, it
can be shown that the Dirac propagator in $d+\frac{1}{2}$
dimensions, $S_F$, contains only one reflection. To write $S_F$
explicitly, it is convenient to take into account the translation
invariance with respect to the first $d$ coordinates $x_\perp =
(x_0, x_1, \ldots , x_{d-1})$, by using a mixed Fourier
transformation: denoting the $x$ coordinates in ${\mathbb
R}^{d+1}$ by $x = (x_\perp ; s)$ , while $s \equiv x_d$, the
`mixed' Fourier transform ${\tilde f}$ of a function $f(x_\perp ;
s)$ will be determined by,
\begin{equation}\label{eq:deffourier}
f(x_\perp;s) \;=\; \int \frac{d^dp}{(2 \pi)^d}  \; e^{i p \cdot
x_\perp } \; {\tilde f}(p;s)   \;,
\end{equation}
where $p$ will denote a momentum vector (we shall omit writing the
label $\perp$ for $d$-dimensional vectors like $p$ wherever there
is no risk of confusion).

Then the explicit solution for ${\tilde S}_F$ is:
\begin{equation}
{\tilde S}_F(p;s,s') \;=\; {\tilde S}_F^{(0)}(p;s,s') \,+\,
{\tilde S}_F^{(1)}(p;s,s') \;,
\end{equation}
where,
\begin{equation}
{\tilde S}_F^{(0)} (p; s,s') \;=\; \frac{1}{2} \big[ \gamma_d \,
{\rm sgn}(s - s') - i \not \! {\hat p} \big] \; e^{ - |p| \, | s -
s' | }
\end{equation}
and
\begin{eqnarray}
{\tilde S}_F^{(1)} (p; s,s') &=& - {\tilde S}_F^{(0)} (p; s, -s')
\,
\gamma_d   \nonumber\\
&=& - \frac{1}{2} \big[ {\rm sgn}(s + s') + i \gamma_d \, \not \!
{\hat p} \big] \; e^{ - |p | \, | s + s' | }.
\label{SF1}
\end{eqnarray}
In the above equation `${\rm sgn}$' stands for the `sign'
function, $|p|$ is the modulus of $p_\mu$, and ${\hat p}_\mu
\equiv \frac{p_\mu}{|p|}$. Notice that $s,s'> 0$ in ${\mathcal
M}$, thus we may ignore the \mbox{${\rm sgn}(s+s')$} in ${\tilde
S}_F^{(1)}$, as that function is identically equal to $1$ there.

Equipped with the expression for ${\tilde S}_F$, it is reassuring
to check that the equations (\ref{eq:deffer}) are indeed
satisfied. In the mixed Fourier representation, the first equation
is tantamount to:
\begin{equation}\label{eq:verif1}
\big( \gamma_d \partial_s + i \not \! {\hat p} \big) {\tilde S}_F
(p;s,s') \;=\; \delta (s - s') \;.
\end{equation}
Then it is immediate to see that
\begin{eqnarray}
\big( \gamma_d \partial_s + i \not \! {\hat p}\big) {\tilde
S}_0 (p;s,s') &=& \delta (s - s') \nonumber\\
\big( \gamma_d \partial_s + i \not \! {\hat p}\big) {\tilde S}_1
(p;s,s') &=& - \,\delta (s + s') \,\gamma_d \;=\; 0 \;,
\end{eqnarray}
where in the last equation we used the fact that $s$ and $s'$ are
in ${\mathcal M}$, thus $\delta(s+s')=0$.

Regarding the bag boundary conditions, using the notation
\mbox{${\mathcal P}_L \equiv \frac{1}{2} (1 - \gamma_d)$}, we see
that:
\begin{eqnarray}
\lim_{s \to 0+} {\mathcal P}_L {\tilde S}_F (p;s,s') &=&- \,
\frac{1}{2} \lim_{s \to 0+} {\mathcal P}_L \big[ {\rm sgn}(s-s')
\,
e^{- |p| \,  |s'|} + {\rm sgn}(s+s') e^{- |p| \, |s'|} \big] \nonumber\\
&=& 0\;\;\;\;, \;\;\;\; \forall s' > 0 \;.
\end{eqnarray}

Let us consider now the calculation of the one-loop vacuum
polarization tensor, $\Pi_{\alpha \beta}$, in ${\mathcal M}$.
Since we have the calculation response to an external constant
electric field in mind as an application, a small momentum
expansion is sufficient.

Adopting the convention that indices from the beginning of the
Greek alphabet ($\alpha,\,\beta, \ldots$) will run from $0$ to
$d$, while the ones from the middle ($\mu, \nu,\ldots$) will
correspond to the range $0,1,\ldots,d-1$, we note that the mixed
Fourier representation for $\Pi_{\alpha \beta}$, denoted ${\tilde
\Pi}_{\alpha \beta}$, is:
\begin{equation}
{\tilde \Pi}_{\alpha \beta} (k;s,s') \;=\; - e^2 \,
\int \frac{d^d p}{(2 \pi)^d} \,
{\rm tr}\Big[ \, {\tilde S}_F(p;s',s) \gamma_\alpha {\tilde S}_F(p+k;s,s')
\gamma_\beta\Big] \;,
\end{equation}
and then we recall the MRE of $S_F$ to obtain,
\begin{equation}
{\tilde \Pi}_{\alpha \beta } (k;s,s') \;=\; \sum_{l,m=0}^1 {\tilde
\Pi}^{(lm)}_{\alpha \beta} (k;s,s'), \label{Pimomentum}
\end{equation}
where,
\begin{equation}
{\tilde \Pi}^{(lm)}_{\alpha \beta} (k;s,s') \;=\; - e^2 \,
 \int \frac{d^d p}{(2 \pi)^d} \,
{\rm tr}\Big[ \, {\tilde S}^{(l)}_F(p;s',s) \gamma_\alpha {\tilde
  S}^{(m)}_F(p+k;s,s') \gamma_\beta \Big] \;.
\label{Pigeral}
\end{equation}
The different terms in ${\widetilde{\Pi}}_{\alpha \beta}$ are then
obtained by evaluating the previous expression for the
corresponding values of $l$ and $m$.

To begin with, we note that ${\tilde\Pi}^{(00)}_{\alpha \beta}$
coincides with the vacuum polarization function in the absence of
boundaries, or, what is equivalent, with boundaries at infinity.
Then, to determine the contributions which truly come from the
boundary, we should consider the remaining components ${\tilde
\Pi}^{(lm)}_{\alpha \beta}$ with at least one of the indices $l,m$
different from $0$.

So, let us first consider the terms ${\tilde \Pi}^{(01)}$ and  ${\tilde
  \Pi}^{(10)}$:
$$
{\tilde \Pi}^{(01)}_{\alpha \beta}(k;s,s') \;=\; \frac{e^2}{4} \,
 \int \frac{d^d p}{(2 \pi)^d} \,
{\rm tr}\Big[ \,\big(\gamma_d \, {\rm sgn}(s' - s) - i \not \! {\hat p}  \big)
\gamma_\alpha \big( 1 + i \gamma_d \,  \gamma\cdot\widehat{(p + k)}\big)
 \gamma_\beta \Big]\;,
$$
\begin{equation}
\times \; \exp[- |p| \, |s - s'| - | p + k | \, |s + s'|]
\label{Pi01}
\end{equation}
$$
{\tilde \Pi}^{(10)}_{\alpha \beta}(k;s,s') \;=\; \frac{e^2}{4} \,
 \int \frac{d^d p}{(2 \pi)^d} \,
{\rm tr}\Big[ \,\big( 1 + i \gamma_d \not \! {\hat p}  \big)
\gamma_\alpha \big( \gamma_d \, {\rm sgn}(s - s') - i
\gamma\cdot\widehat{(p + k)}\big)
 \gamma_\beta \Big]
$$
\begin{equation}
\times \; \exp[- |p| \, |s + s'| - | p + k | \, |s - s'|] \;,
\label{Pi10}
\end{equation}
where $(\widehat {p+k})_{\mu}\equiv (p+k)_{\mu}/|p+k|$.

It is immediate to notice that the results for ${\tilde \Pi}^{(01)}$ and
${\tilde \Pi}^{(10)}$ depend on the spacetime dimension ($d+1$) being
even or odd. Indeed, both contributions vanish when the spacetime
dimension is even, since they involve the trace of an odd number (3
and 5) of $\gamma$ matrices. Even for the case of an odd $(d+1)$, those
traces can only be different from zero for $d=2$ and $d=4$. Namely,
there is no parity breaking (to one-loop order) for any $d>4$ (even or
odd).

In what follows, when considering these contributions, we shall assume
that $d=2$, a physically interesting situation, since it corresponds
to planar physics in a system with a border. This is potentially
interesting when studying effective condensed matter physics models
with Dirac fermions in $2+1$ dimensions coupled to a gauge field and
confined to a half-plane.

Besides, notice that ${\tilde \Pi}^{(01)}$ and ${\tilde \Pi}^{(10)}$ yield
parity-breaking contributions, since the trace of an odd number of
matrices will necessarily introduce, in $2+1$ dimensions, a
Levi-Civita tensor~\footnote{That will not be so for the ${\tilde
    \Pi}^{(11)}$ term, which, as we shall see, is parity-conserving and
  different from zero for all $d$.}. This parity-breaking radiative effect
is due to the explicit breaking of parity by the boundary
condition~\cite{Deser:1981wh}.

Combining the two contributions, ${\tilde \Pi}^{(01)}$ and ${\tilde
  \Pi}^{(10)}$, into a single ${\tilde \Pi}^{odd} \equiv {\tilde
  \Pi}^{(01)}+{\tilde \Pi}^{(10)}$ tensor, we see (after taking the
traces and performing a change of variables) that its $\mu, \nu$
components are given by:
\begin{eqnarray}
{\tilde \Pi}^{odd}_{\mu\nu}(k;s,s') &=& i \frac{e^2}{2} \,
\int \frac{d^2 p}{(2 \pi)^2}\;
\Big\{ \big[ \varepsilon_{\mu\nu} \, {\rm sgn}(s' - s) \, + \,
\frac{\varepsilon_{\gamma \mu} p_\gamma k_\beta +
\varepsilon_{\gamma \nu }
 p_\gamma k_\alpha}{|p| \; |p + k|} \nonumber\\
&-& \delta_{\mu \nu} \, \frac{ p \land k}{|p| \; |p + k|} \big] \;
\exp[- |p| \, |s - s'| - | p + k | \, |s + s'|]    \nonumber\\
 &+& \big[ \varepsilon_{\mu\nu} \, {\rm sgn}(s' - s) \, - \,
\frac{\varepsilon_{\gamma \mu} p_\gamma k_\nu +
\varepsilon_{\gamma \nu }
 p_\gamma k_\mu}{|p| \; |p - k|} \nonumber\\
&+& \delta_{\mu \nu} \, \frac{ p \land k}{|p| \; |p - k|} \big] \;
\exp[- |p| \, |s - s'| - | p - k | \, |s + s'|] \Big\}   \;,
\label{Piodd}
\end{eqnarray}
where we have introduced the notation $p \land q \equiv \varepsilon_{\mu\nu} p_\mu q_\nu$.  This
is still a rather complicated expression to deal with. However, for
the response to a constant field it is enough to perform an expansion
in powers of $k$, and evaluate the different contributions to ${\tilde
  \Pi}^{odd}$.  The leading term when $k \to 0$ (required for the
calculation of the induced current) is:
\begin{equation}
{\tilde \Pi}^{odd}_{\mu\nu}(0;s,s') \;=\; -i \frac{ e^2}{2 \pi} \, \frac{ {\rm
    sgn}(s' - s) }{\big(|s - s'| + |s + s'|\big)^2 } \varepsilon_{\mu \nu} \;.
\label{Piodd1}
\end{equation}
There are, of course, also components: ${\tilde \Pi}^{odd}_{dd}(k;s,s')$
and ${\tilde \Pi}^{odd}_{d\mu}(k;s,s')$.  However, they vanish identically
when $k \to 0$.

Let us now calculate the remaining term ${\tilde \Pi}^{(11)}_{\alpha \beta}$ in
$d+1$ spacetime dimensions. Using (\ref{Pigeral}) and (\ref{SF1}) we
see that it is given by:
$$
{\tilde \Pi}^{(11)}_{\alpha \beta}(k;s,s') \;=\;-\frac{e^2}{4} \,
\int \frac{d^d p}{(2 \pi)^d} \,
{\rm tr}\Big[ \,\big(1 - i\gamma_d \not \! {\hat p}  \big)
\gamma_\alpha \big( 1 + i \gamma_d \widehat{( \not \! {p} +
\not \! {k})}\big)
\gamma_\beta \Big]
$$
\begin{equation}
\times \; \exp[- (|p|+| p + k |)|s + s'|] \;,
\label{Pi11ab}
\end{equation}
where $s,\,s^\prime > 0$.
To perform the trace operations in (\ref{Pi11ab}), we consider the
cases ${\tilde \Pi}^{(11)}_{\mu \nu}$, ${\tilde \Pi}^{(11)}_{d \mu}$ and
${\tilde \Pi}^{(11)}_{d d}$ separately.  We get,
$$
{\tilde \Pi}^{(11)}_{\mu \nu}(k;s,s') \;=\;-\frac{e^2}{4} r\,
\int \frac{d^d p}{(2 \pi)^d} \,
 \exp[- (|p|+| p + k |)|s + s'|]
$$
\begin{equation}
\times \; \Big[ \delta_{\mu \nu} - \hat {p}_{\mu}(\widehat {p+k})_{\nu} -
\hat {p}_{\nu}(\widehat {p+k})_{\mu} +
\hat {p}\cdot (\widehat {p+k})\delta_{\mu \nu} \Big]],
\label{Pi11munu}
\end{equation}
where $r$ stands for the trace of the unit matrix in a
($d+1$)-dimensional spacetime.  In the $k \to 0$ limit,
\begin{equation}
{\tilde \Pi}^{(11)}_{\mu \nu}(0;s,s') \;=\;-\frac{e^2}{2} r\,
\left( 1-\frac{1}{d} \right)\delta_{\mu \nu} I_{0}(s+s'),
\label{Pi11munuzero}
\end{equation}
where $I_{0}(s+s')$ may be obtained from the general integral,
\begin{equation}
I_{n}(t)=\int \frac{d^d p}{(2 \pi)^d} \, \frac{\exp[- 2|p||t|]}{|p|^n}
=
\frac{2^{1-2d+n}\Gamma(d-n)}{\pi^{d/2}\Gamma(d/2)}\frac{1}{|t|^{d-n}}.
\label{Inss}
\end{equation}

For the ${\tilde \Pi}^{(11)}_{d \mu}$ component, we obtain from
(\ref{Pi11ab}) the expression,
\begin{equation}
{\tilde \Pi}^{(11)}_{d \mu}(k;s,s') \;=\;-i\frac{e^2}{4} r\,
\int \frac{d^d p}{(2 \pi)^d} \,
 \exp[- (|p|+| p + k |)|s + s'|]\left[(\widehat {p+k})_\mu -
\hat{p}_\mu \right] \;,
\label{Pi11dmu}
\end{equation}
which leads to ${\tilde \Pi}^{(11)}_{d \mu}(0;s,s') \;=\;0$.

Analogously, for ${\tilde \Pi}^{(11)}_{d d}$:
\begin{equation}
{\tilde \Pi}^{(11)}_{d d}(k;s,s') \;=\;-\frac{e^2}{4} r\,
\int \frac{d^d p}{(2 \pi)^d} \,
 \exp[- (|p|+| p + k |)|s + s'|]\left[1-\hat{p}\cdot(\widehat
   {p+k})\right] \;,
\label{Pi11dd}
\end{equation}
and also in this case we have ${\tilde \Pi}^{(11)}_{d d}(0;s,s') \;=\;0$.

The small-$k$ expressions for ${\tilde \Pi}^{(11)}_{d d}$ and ${\tilde
  \Pi}^{(11)}_{d \mu}$ are:
\begin{equation}
{\tilde \Pi}^{(11)}_{d \mu}(k;s,s') \;=\;-i\frac{e^2}{4} r\,
\left(1-\frac{1}{d} \right)k_{\mu}\,I_{1}(s+s'),
\label{Pi11dmu}
\end{equation}
and
\begin{equation}
{\tilde \Pi}^{(11)}_{d d}(k;s,s') \;=\;-\frac{e^2}{4} r\,
\left[-2|s+s'|\frac{1}{d}k^{2}\,
I_{1}(s+s') +\frac{1}{2}k^{2}\left(1-\frac{5}{d}\right)\,
I_{2}(s+s') \right].
\label{Pi11dd}
\end{equation}
where $I_{1}(s+s')$ and $I_{2}(s+s')$ are obtained from (\ref{Inss}).
One can also verify that for small $k$,
\begin{equation}
{\tilde \Pi}^{(00)}_{d d}(k;s,s') \;=\;-{\tilde \Pi}^{(11)}_{d d}(k;s,-s')
\end{equation}
and
\begin{equation}
{\tilde \Pi}^{(00)}_{d \mu}(k;s,s') \;=\;
-{\tilde \Pi}^{(11)}_{d \mu}(k;s,-s') {\rm sgn}(s'-s).
\end{equation}

In the linear response approximation, the induced current can be
written as follows:
\begin{equation}
j_{\alpha}(x_{\perp},s) = -i \int dy_{\perp} ds' \Pi_{\alpha \beta}(x_{\perp},
y_{\perp};s,s')A_{\beta}(y_{\perp},s') \,,
\label{current1}
\end{equation}
where we follow the notation introduced in (\ref{eq:deffourier}).  We
are interested in the particular case of a constant electric field
with modulus $E$ and normal to the wall. In the gauge $A_{\alpha}=0\,,\;
\alpha=1,...,d$ we have,
\begin{equation}
A_0(y_{\perp},s')\;=\;-\, E \;  s'
\end{equation}
and
\begin{equation}\label{current2}
j_{\alpha}(x_{\perp},s) \;=\; i \,E \; \int dy^{0}_{\perp}
\int d^{d-1}y_{\perp} \int^{\infty}_{0}ds'
\Pi_{\alpha 0}\Big(x_{\perp}- y_{\perp};s,s'\Big) \; s' \;.
\end{equation}
Translation invariance in $x_\perp$ implies of course that the current is
independent of $x_{\perp}$, so $j_\alpha = j_\alpha(s)$ (as can be seen by a shift
of variables).

Next we distinguish the cases of spacetime dimension $d+1$ even or odd
($d=2$).  In the first case, since ${\tilde \Pi}^{odd}= {\tilde
  \Pi}^{(01)}+{\tilde \Pi}^{(10)}=0$, and both ${\tilde \Pi}^{(11)}_{\mu \nu}$
and ${\tilde \Pi}^{(00)}_{\mu \nu}$ are proportional to $\delta_{\mu \nu}$, only the
$j_{0}$ component is different from zero. On the other hand, when
$d=2$, both $j_0$ and $j_1$ may be non-vanishing, since ${\tilde
  \Pi}^{odd}_{l 0}(0;s,s')$ is different from $0$. For $j_{0}$ (and any
$d$), the contribution always comes only from ${\tilde \Pi}^{(11)}_{0
  0}(0;s,s')$.

Now, recalling (\ref{Piodd1}), we obtain for $d=2$
\begin{equation}
j_1(s) \;=\; - \, E \frac{ e^2}{2 \pi} \,\int^{\infty}_{0} ds'
\frac{ {\rm sgn}(s' - s) }{\big(|s - s'| + |s + s'|\big)^2 } \;,
\label{jodd1}
\end{equation}
and evaluating the integral over $s'$:
\begin{equation}
j_{1}(s)\;=\;- \,\frac{e^{2} E}{4\pi s}\, .
\label{j1}
\end{equation}
This result shows that there is indeed a non-trivial parity breaking
effect due to the boundary: the induction of a current {\em
  transverse\/} to the electric field. That current is also
concentrated on the wall ($s=0$), where parity is maximally violated.

Analogously, from (\ref{Pi11munuzero}), and introducing an ultraviolet
regularization through a small $\epsilon$ parameter, we obtain,
$$
j_{0}(s)= - i \frac{e^{2}E r(d-1)}{2 d}
\frac{2^{1-2d}\Gamma(d)}{\pi^{d/2}\Gamma(d/2)}$$
\begin{equation}
\times
\left\{ \int^{\infty}_{0} s'ds'\frac{1}{\left[(s-s')^{2}+\epsilon^{2}
\right]^{d}}+
\int^{\infty}_{0} s'ds'\frac{1}{\left[(s+s')^{2}+\epsilon^{2}
\right]^{d}}
\right\}\;,
\end{equation}
for any $d$. When $d=2$, IR divergences are present, which we
deal with by means of an IR cutoff $L$. After some standard
calculations, the leading contribution for a small $\epsilon$ is:
\begin{equation}
j_{0}(s)=-i \frac{e^{2} E}{8 \pi}\left[-ln(\frac{s}{L})+
\frac{\pi s}{2 \epsilon}\right] \;.
\label{j0}
\end{equation}

We see from (\ref{j1}) and (\ref{j0}) that the $j_{1}(s)$ component is
divergent only one the boundary ($s=0$). The $j_{0}(s)$ component, on
the other hand, has a space-dependent UV divergence which vanishes on
the wall.

Let mention that the Ward identity for $\Pi_{\alpha\beta}$, not
evident in the different expressions we have obtained for
$\Pi_{\alpha\beta}$, can be checked explicitly.

Finally, we mention that, had the boundary been located at an
arbitrary point $s^0$ rather than at $s=0$, the resulting
${\tilde\Pi}_{\alpha\beta}$ could have been easily obtained from
the one for $s=0$ by the change of variables: $s \to s - s^0$ and
$s' \to s' - s^0$ in ${\tilde\Pi}_{\alpha\beta}$. Namely:
\begin{equation}
\Big[ {\tilde\Pi}_{\alpha\beta}(k;s,s')\Big]_{s_0\neq 0} \;=\;
\Big[ {\tilde\Pi}_{\alpha\beta}(k;s-s^0,s'-s^0) \Big]_{s_0 = 0}\;.
\end{equation}
The reason is that the fermion propagator in the new variables may
be obtained by the same change of variables. That this is so can
be checked, for example, by noting that both the differential
equation and the bag condition~\ref{eq:deffer} are then
automatically satisfied on the new boundary.

It is then quite easy to obtain consider, for example, the
expressions for the vacuum polarization function in the limit of a
``wall at infinity´´, which corresponds to $s^0 \to - \infty$. One
easily verifies that, for $s, s' > s^0$, all the terms
${\tilde\Pi}^{(l m)}$ with $l$ or $m$ different from $0$ vanish.
The result tends to the usual, no-reflection one.

\section*{Acknowledgments}
C.D.F. was partially supported by a Fundaci\'on Antorchas grant. The
authors acknowledge CAPES, CNPq and Pronex/MCT for partial financial support.


\begin{thebibliography}{99}
\bibitem{Casimir:1948dh} H.~B.~G.~Casimir,
Kon.\ Ned.\ Akad.\ Wetensch.\ Proc.\  {\bf 51}, 793 (1948).
\bibitem{Bordag:2001qi} See, for example:\\
M.~Bordag, U.~Mohideen and V.~M.~Mostepanenko,
Phys.\ Rept.\ {\bf 353}, 1 (2001), for a modern review and current
status of the problem.
\bibitem{Chodos:1974pn}
A.~Chodos, R.~L.~Jaffe, K.~Johnson and C.~B.~Thorn,
Phys.\ Rev.\ D {\bf 10}, 2599 (1974).
\bibitem{Fuentes:1995ym} See, for example:\\
M.~Fuentes, A.~Lopez, E.~H.~Fradkin and E.~Moreno,
Nucl.\ Phys.\ B {\bf 450}, 603 (1995).
\bibitem{Deser:1981wh}
For a throughout study of the interplay between symmetries and
radiative effects in $2+1$ dimensional topoloically massive $QED$ see:\\
 S.~Deser, R.~Jackiw and S.~Templeton,
Annals Phys.\  {\bf 140}, 372 (1982)
[Erratum-ibid.\  {\bf 185}, 406.1988\ APNYA,281,409 (1988\ APNYA,281,409-449.2000)].
\bibitem{BB1} R. Balian and C. Bloch , Ann. Phys. (N.Y.) {\bf 60}, 401 (1970).
\bibitem{BB2} R. Balian and C. Bloch , Ann. Phys. (N.Y.) {\bf 64}, 271 (1971).
\bibitem{HJ1} T. H. Hanson and R. L. Jaffe, Phys. Rev. D {\bf 28}, 882 (1983).
\bibitem{HJ2} T. H. Hanson and R. L. Jaffe, Ann. Phys. (N.Y.) {\bf 151}, 204
(1983).
\bibitem{K} J. I. Kapusta, Finite Temperature Field Theory (Cambridge
University Press, New York, 1989).
\bibitem{ZJ} J. Zinn-Justin, Quantum Field Theory and Critical Phenomena
(Clarendon Press, Oxford, 1996).
\end{thebibliography}
\end{document}